\documentclass[useAMS,usenatbib]{mn2e}

\usepackage{times}

\usepackage[dvips]{graphics}

\title[Dynamically active clouds]
{Dense and diffuse gas in dynamically active clouds}
\author[R. T. Garrod, D. A. Williams, J. M. C. Rawlings]
    {R. T. Garrod,$^{1,2}$\thanks{E-mail:rgarrod@mps.ohio-state.edu}
    D. A. Williams$^{2}$ and J. M. C. Rawlings$^{2}$\\
    $^{1}$Department of Physics,
    Ohio State University, 191 W. Woodruff Avenue, Columbus, OH 43210, USA\\
    $^{2}$Department of Physics and Astronomy, 
    University College London, Gower Street, 
    London WC1E 6BT, UK}

\begin{document}

\date{Accepted . Received ; in }

\pagerange{\pageref{firstpage}--\pageref{lastpage}} \pubyear{2006}

\maketitle

\label{firstpage}

\begin{abstract}

We investigate the chemical and observational implications of repetitive transient dense core formation in molecular clouds. We allow a transient density fluctuation to form and disperse over a period of 1 Myr, tracing its chemical evolution. We then allow the same gas immediately to undergo further such formation and dispersion cycles. The chemistry of the dense gas in subsequent cycles is similar to that of the first, and a limit cycle is reached quickly (2 -- 3 cycles). Enhancement of hydrocarbon abundances during a specific period of evolution is the strongest indicator of previous dynamical history. The molecular content of the diffuse background gas in the molecular cloud is expected to be strongly enhanced by the core formation and dispersion process. Such enhancement may remain for as long as 0.5 Myr. The frequency of repetitive core formation should strongly determine the level of background molecular enhancement.

We also convolve the emission from a synthesised dark cloud, comprised of ensembles of transient dense cores. We find that the dynamical history of the gas, and therefore the chemical state of the diffuse inter-core medium, may be determined if a sufficient sample of cores is present in an ensemble. Molecular ratios of key hydrocarbons with SO and SO$_2$ are crucial to this distinction. Only surveys with great enough angular resolution to resolve individual cores, or very small groupings, are expected to show evidence of repetitive dynamical processing. The existence of non-equilibrium chemistry in the diffuse background may have implications for the initial conditions used in chemical models. Observed variations in the chemistries of diffuse and translucent regions may be explained by lines of sight which intersect a number of molecular cloud cores in various stages of evolution.

\end{abstract}


\begin{figure*}
\centering
\scalebox{0.55}[0.7]{\includegraphics{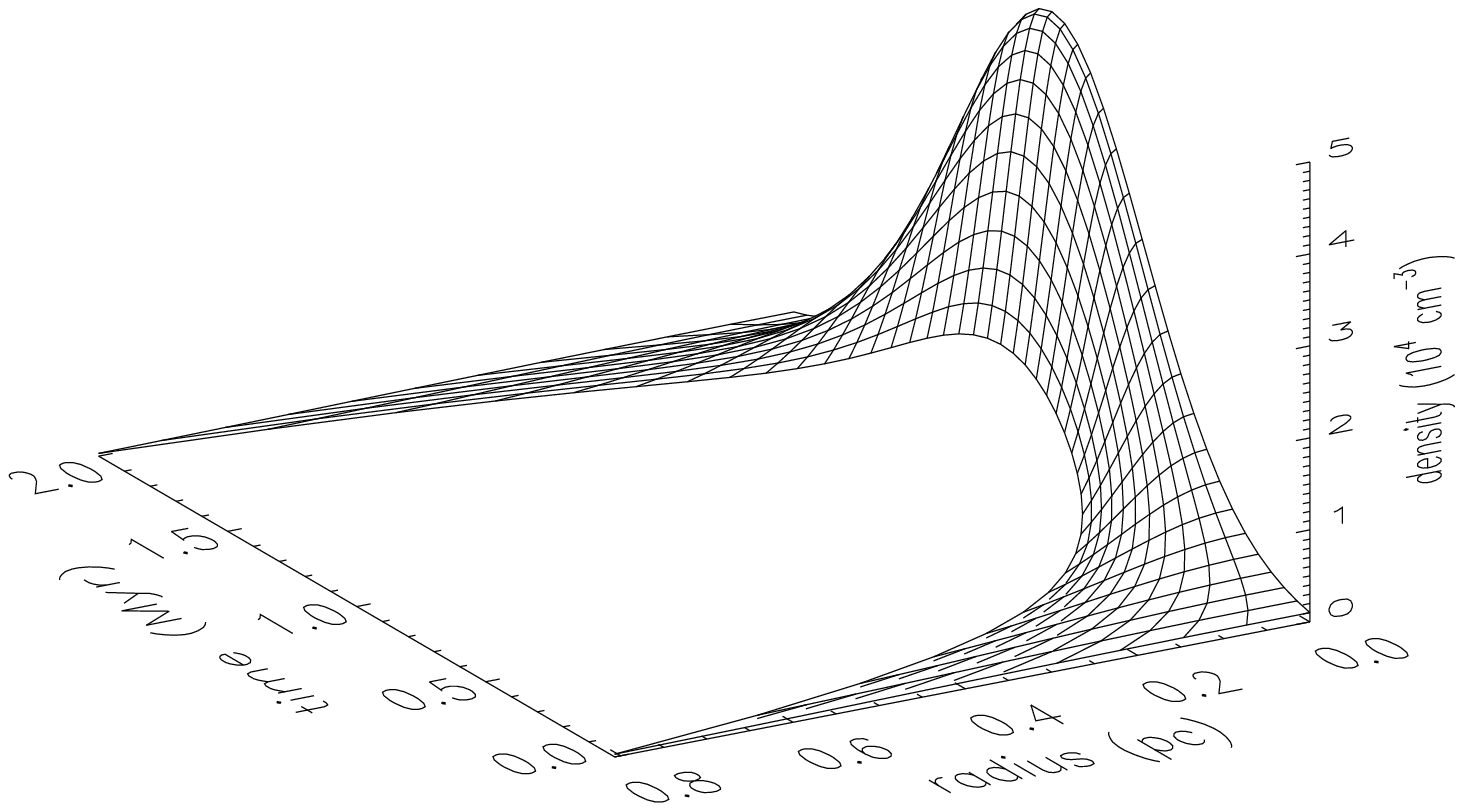}}
\scalebox{0.55}[0.7]{\includegraphics{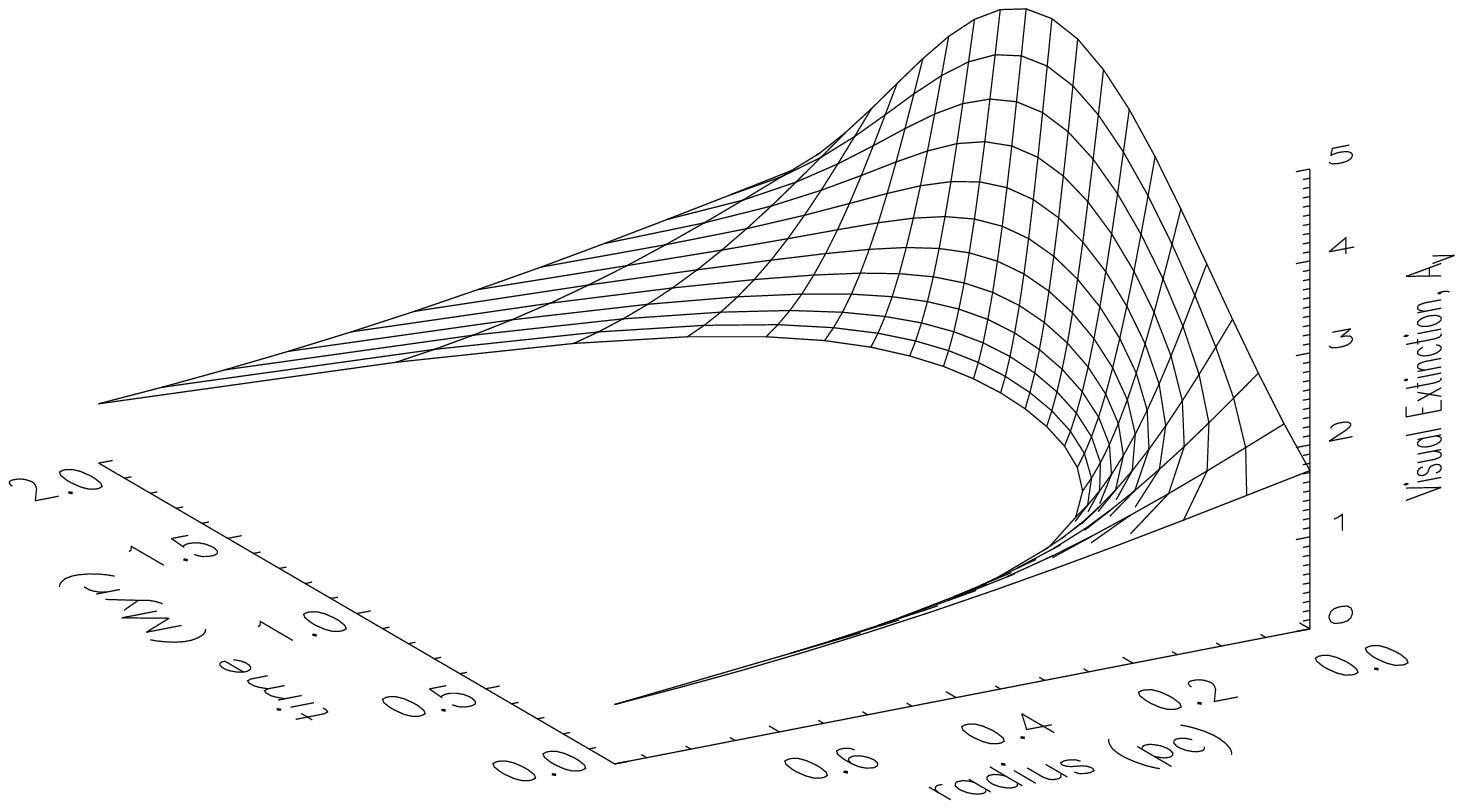}}
\caption{\label{fig0} Time-evolution of density and visual extinction profiles along a radius on the principal collapse axis.}
\end{figure*}

\section{Introduction}

Molecular line observations at high angular resolution of two dark interstellar cloud regions -- L673 and TMC-1, core D -- indicate that their emissions are dominated by small-scale clumpy structure, previously unresolved \cite[]{morata03a,morata05a,peng98a}. The structure is required by the observations to be transient. Such dynamical behaviour in an interstellar cloud will have profound effects on the chemistry of the molecular tracers by which the dense gas in the cloud is mapped, and the observed clumpiness may also have consequences for low mass star formation. 

In a previous paper \nocite{garrod06a} (Garrod et al. 2006, hereafter GWR) we studied the chemistry of a model cloud assumed to be composed of a large number of small, transient dense cores randomly distributed in a low density background; the cores are at random evolutionary phases. We showed that such a model cloud exhibits very successfully the generally observed characteristics of interstellar cloud maps, at both high and low resolutions, in terms of sizes of emitting regions for different species, their column densities, the separation of their emission peaks, and also ice mantle composition and growth.

The transience of the dense cores implies that gas in interstellar clouds may be re-cycled between high and low density phases several times within the lifetime of the cloud. \nocite{garrod05a} Garrod et al. (2005, hereafter GWHRV) have shown that, for a particular core, a single cycle from low to high and back to low density gives rise to a chemical hysteresis, in which the final chemical state differs from the initial state. In principle, if cycle number could be determined from the chemistry, it would be a useful indicator of evolutionary status.

The purpose of the present paper is to explore the chemical consequences of re-cycling gas through several dynamical cycles. There are two main aims:

\begin{enumerate}
\item We shall examine the consequences of re-cycling on the chemistry of the low density background gas. Are there signatures in low density gas of repeated successive cycling to high density? Is the chemistry of the low density gas significantly altered by such re-cycling? If so, can the observed chemical variety of different diffuse clouds be accounted for by this process?

\item We shall also consider whether the enrichment in the background gas from one cycle affects the chemistry in a dense core in succeeding cycles. If tracers of re-cycling do exist in one dense core, are these effects detectable in ensembles of cores with a range of evolutionary ages?
\end{enumerate}

In Section 2 we describe the model cloud used in these studies, and investigate the chemical effects on the dense gas of repetitive dynamical processing. Section 3 deals with results for the diffuse background. In section 4 we explore the observational effects of dynamical processing. We present our conclusions in Section 5.

\begin{table*}
\renewcommand{\thefootnote}{\fnsymbol{footnote}}
\centering
\begin{minipage}{105mm}
\caption{\label{tab1} Selected column densities through the core and fractional abundances computed from column densities, for three cycles at hydrocarbon ``shoulder'' (0.72 Myr). Values which vary from the previous cycle by more than a factor of 2 are shown in bold face.$^{a}$}
\begin{tabular}{|l|c|c|c|c|c|c|c|c|c|}
\hline
Species $i$ & & \multicolumn{2}{c}{Cycle 1} & &
\multicolumn{2}{c}{Cycle 2} & & \multicolumn{2}{c}{Cycle 3} \\
 & & $N[i]$ & $X_{N}[i]$ $^{b}$ & &
$N[i]$ & $X_{N}[i]$ $^{b}$ & &
$N[i]$ & $X_{N}[i]$ $^{b}$ \\
 & & (cm$^{-2}$) &  &  &
(cm$^{-2}$) &  & &
(cm$^{-2}$) &  \\
\hline
CO & & 7.9(17) & 7.0(-05)  &
& 8.3(17) & 7.4(-05)  &
& 8.3(17) & 7.4(-05) \\

O$_2$ & & 6.1(14) & 5.4(-08)  &
& {\bfseries 1.2(15)} & {\bfseries 1.1(-07)}  &
& 1.2(15) & 1.1(-07) \\

H$_2$O & & 2.1(15) & 1.9(-07)  &
& 1.8(15) & 1.6(-07)  &
& 1.8(15) & 1.6(-07) \\

OH & & 1.4(14) & 1.2(-08)  &
& 1.2(14) & 1.1(-08)  &
& 1.2(14) & 1.1(-08) \\

CH & & 1.3(14) & 1.2(-08)  &
& 1.2(14) & 1.1(-08)  &
& 1.2(14) & 1.1(-08) \\

HCO$^+$ & & 6.0(12) & 5.3(-10)  &
& 3.7(12) & 3.3(-10)  &
& 3.6(12) & 3.3(-10) \\

H$_2$CO & & 1.4(14) & 1.2(-08)  &
& {\bfseries 5.1(13)} & {\bfseries 4.5(-09)} &
& 5.0(13) & 4.4(-09) \\

CH$_3$OH & & 4.1(12) & 3.6(-10)  &
& {\bfseries 9.6(11)} & {\bfseries 8.5(-11)}  &
& 9.1(11) & 8.1(-11) \\

NH$_3$ & & 7.2(12) & 6.4(-10)  &
& 5.2(12) & 4.6(-10)  &
& 5.1(12) & 4.5(-10) \\

N$_{2}$H$^{+}$ & & 1.6(11) & 1.4(-11)  &
& 1.0(11) & 9.0(-12)  &
& 9.8(10) & 8.7(-12) \\

CN & & 3.2(13) & 2.8(-09)  &
& 1.2(13) & 1.1(-09) &
& 1.2(13) & 1.1(-09) \\

NO & & 2.3(14) & 2.1(-08)  &
& 3.2(14) & 2.8(-08)  &
& 3.2(14) & 2.8(-08) \\

CS & & 4.5(14) & 4.0(-08)  &
& 3.5(14) & 3.1(-08)  &
& 3.5(14) & 3.1(-08) \\

SO & & 9.3(12) & 8.2(-10)  &
& {\bfseries 1.9(13)} & {\bfseries 1.7(-09)}  &
& 2.0(13) & 1.8(-09) \\

SO$_2$ & & 1.9(11) & 1.7(-11)  &
& {\bfseries 9.7(11)} & {\bfseries 8.6(-11)}  &
& 1.1(12) & 9.4(-11) \\

CH$_4$ & & 7.3(14) & 6.5(-08)  &
& {\bfseries 1.3(14)} & {\bfseries 1.2(-08)}  &
& 1.2(14) & 1.1(-08) \\

C$_3$H & & 1.1(14) & 9.9(-09)  &
& {\bfseries 1.2(13)} & {\bfseries 1.1(-09)}  &
& 1.1(13) & 9.7(-09) \\

C$_5$H & & 3.0(11) & 2.7(-11)  &
& {\bfseries 1.8(10)} & {\bfseries 1.6(-12)}  &
& 1.7(10) & 1.5(-12) \\

C$_3$H$_4$ & & 1.8(10) & 1.6(-12)  &
& {\bfseries 4.0(08)} & {\bfseries 3.6(-14)}  &
& 3.3(08) & 2.9(-14) \\

HC$_3$N & & 7.2(11) & 6.4(-11)  &
& {\bfseries 1.2(11)} & {\bfseries 1.0(-11)}  &
& 1.0(11) & 9.2(-12) \\
\hline
\end{tabular}
$^{a}$$A(B)=A \times 10^{B}$ \\
$^{b}$$X_{N}[i]=N[i]/(2N[\mbox{H}_{2}]+N[\mbox{H}])$ \\
\end{minipage}
\end{table*}

\section{Cyclic Dense Core Model -- The Dense Gas}

The model of interstellar molecular clouds that we explore in this series of papers is as follows:

We consider a molecular cloud to be an ensemble of units that evolve in time both chemically and physically, in a cyclic manner.

As described in Paper 1 (GWHRV), a unit (transient core) is assumed to grow from a large weak perturbation in a diffuse background gas to a small high density structure which then decays back to low density. The chemistry within this unit responds to these changing physical conditions in a time-dependent way.

In the ensemble which is to represent the interstellar cloud, these units are assumed to be positioned spatially at random within the cloud, and to have random evolutionary phase. As shown in Paper 2 (GWR), column density and line intensity maps may then be computed for the ensemble, synthesising telescope angular resolutions. A line of sight through the cloud may intersect a number of units at different stages of evolution.

This model predicts that a localised region observed to have high intensity in a particular molecular line, commonly called by observers ``a core'', may be a superposition of emission from several units along the line of sight, at different stages of evolution. The convolution of the emission from the separate units, at low angular resolutions, acts to mask the small-scale structure. 

The conclusion of Paper 2 is that such a model, for plausible parameter choices, accounts for the observed structure of molecular clouds, including core sizes, column densities, chemical differentiation in space, and apparent separation between emission peaks.

In this section we utilise the dynamical/chemical model for one unit (transient core), following the effects on the gas chemistry of successive dynamical cycles of formation and dissipation.

In GWHRV, we employed the results of \cite{falle02a} to determine the dynamics of a core, the basic unit within a cloud. Falle \& Hartquist offer a plausible magneto-hydrodynamic mechanism whereby slow-mode MHD waves form large density inhomogeneities as a consequence of the non-linear steepening of fast-mode waves. Guided by their 1-D computations, we constructed space- and time-dependent density and visual extinction profiles, throughout the collapse and subsequent dispersal of a core. The central density of the core varies as a Gaussian function of time, and the density profile along the axis of collapse varies as a Gaussian function with respect to distance from the core centre, normalised to the central density. We use the conservation of mass to constrain the physical size of the core as collapse takes place. We assume that the presence of a magnetic field causes collapse to take place preferentially along one axis, whilst much less vigorous collapse occurs in the other two dimensions. We therefore choose the number of dimensions along which collapse takes place to be $k = 1.5$. To calculate visual extinction, we evaluate the column density of hydrogen along the axis of collapse and multiply by a conversion factor \cite[]{vanDishoeck98a}. The visual extinction value varies during the collapse, since $k > 1$.

In the chemical model, the core is represented by 12 depth points running from centre to edge along the principal axis of collapse. Each point represents the same parcel of gas at all times, and assumes its own time-dependent density and visual extinction as outlined above, and defined below:
\begin{eqnarray}
\hspace{0.1cm}\rho(\alpha,t) & = & \rho(0,t_{m}) \exp \left[ - \alpha ^{2} \right] \exp \left[ - 
\left( \frac{t - t_{m}}{\tau} \right)^{2} \right] \\
\hspace{0.1cm}A_{V}(0,t) & \propto & [\rho(0,t_{m})]^{\frac{1}{k}} [\rho(0,t)]^{(1-\frac{1}{k})}
\end{eqnarray}
In the density expression, $\rho$ is number density, $\alpha$ is the time-independent parametrized distance of the depth point from the core centre, $t$ is time, $\rho(0,t_{m})$ is peak central density, and $t_{m}$ is the time at which peak density is reached (1 Myr); in this paper we refer to a time $t = t_{m}$ as ``peak time''. $\tau$ is chosen according to initial central density. In the visual extinction expression, $k$ is the number of dimensions (1.5) along which collapse and expansion are assumed to take place. Visual extinctions at points outward from the centre adopt a time-independent fraction of the central value. See GWHRV for a full explanation of density and visual extinction profiles.

Figure \ref{fig0} shows the time-evolution of the density and visual extinctions at each depth point. Profiles are shown for every 50,000 years. These profiles also demonstrate the contraction of the core to reach peak density, and its subsequent re-expansion. Initial densities range from $\sim$300 cm$^{-3}$ at the edge of the core to 10$^3$ cm$^{-3}$ at the centre, and grow to $\sim$10$^4$ cm$^{-3}$ and $5 \times 10^4$ cm$^{-3}$ respectively at peak time (1 Myr). Densities and visual extinctions fall back to their initial values over a period of 1 Myr after peak density is achieved. The central visual extinction at peak density time reaches $A_V=5.0$.

The initial run is identical to that explored by GWR. Having run the model through one dynamical cycle, we run the same dynamical/chemical model, using the final fractional abundance values at each depth point as the starting values of the following run. The physical evolution of the core therefore proceeds in the same way for each point as it did in the previous cycle; after the dispersal of the core, the gas immediately undergoes an identical dynamical process. This process is repeated for several more cycles. A more complex model of the cycling process (such that subsequent cycles might process the material in different ways, according to larger or smaller changes to physical parameters, or might be physically displaced from the original site) is beyond the scope of this study.

At each depth point we follow the chemistry of 451 species, using 3631 reactions (including sticking ``reactions'') taken from the UMIST ratefile \cite[]{millar97a}. The initial cycle (the same as that of GWR) is run starting from a chemical steady state for diffuse background conditions. Freeze-out is turned on if and when depth points reach a visual extinction greater than 2.5; immediate re-injection of grain mantles into the gas phase occurs when points fall below this critical value. Observational evidence exists for such a threshold -- see e.g. \cite{whittet01a}. The freeze-out {\em rate} is maintained for all subsequent cycles at the same level as the first, which in fact leads to slightly larger values of freeze-out in later cycles ($\sim$62\% 
of all CO, at peak time, for the second cycle, as compared to 60\% 
for the first).

\begin{figure*}
\centering
\scalebox{0.9}[0.9]{\includegraphics{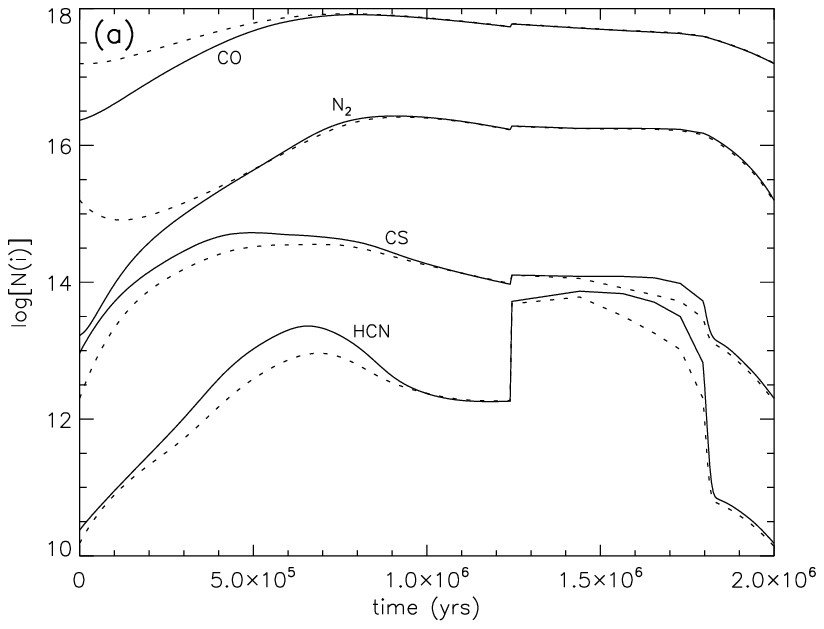}}
\scalebox{0.9}[0.9]{\includegraphics{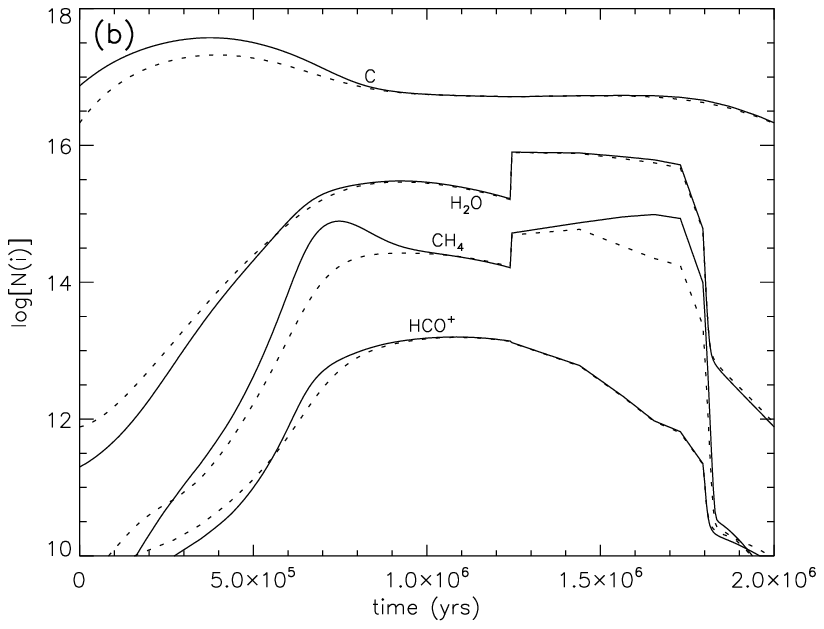}}
\scalebox{0.9}[0.9]{\includegraphics{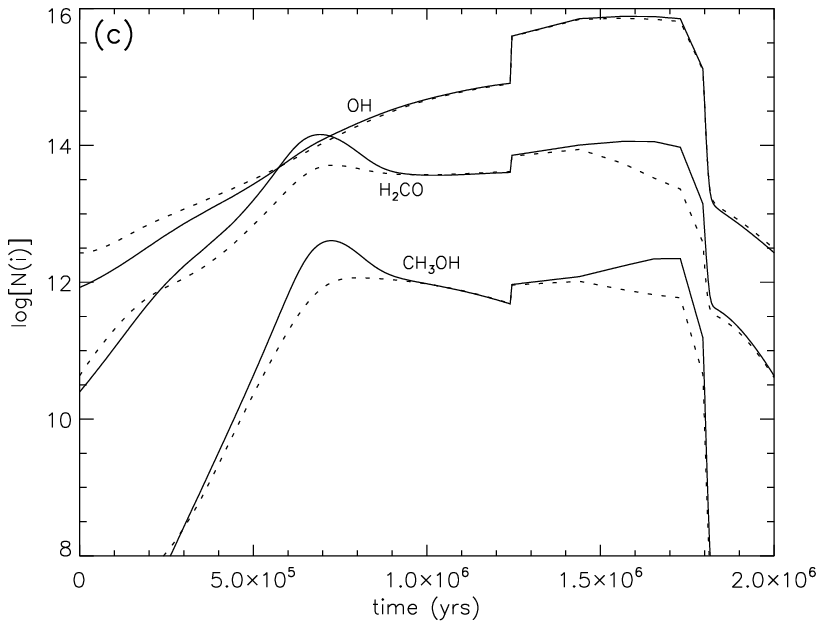}}
\scalebox{0.9}[0.9]{\includegraphics{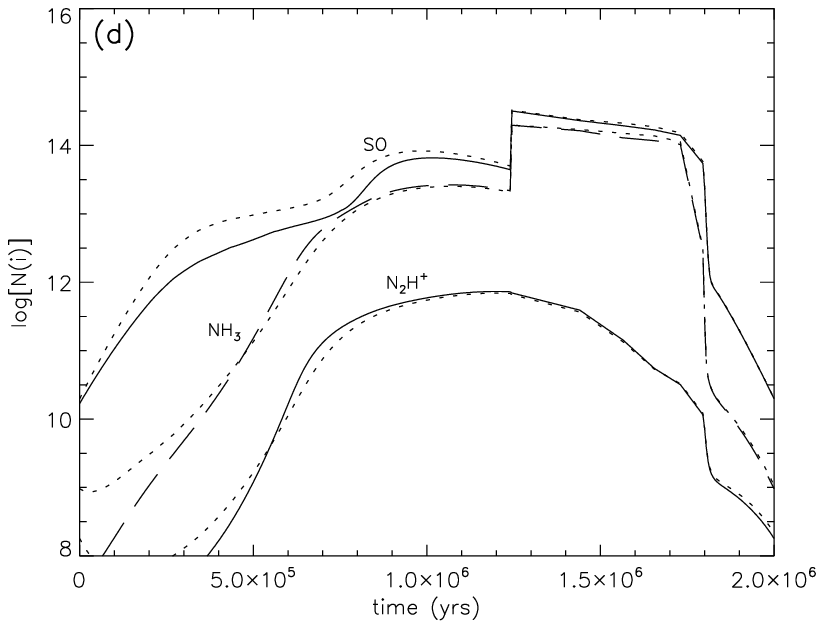}}
\caption{\label{fig1} Column densities as functions of time, calculated through the core -- first cycle (solid lines and dashed lines) and second cycle (dotted lines).}
\end{figure*}

\subsection{Results}

The chemical evolution of the gas has been traced through a number of dynamical cycles, however after the third cycle the differences between abundances at corresponding times in each cycle become negligible or non-existent. This represents the ``limit cycle'' case \cite[see][]{charnley88b}. Figure \ref{fig1} shows column densities for a range of important and/or commonly observed molecules, calculated by integrating the depth-dependent chemical abundances along a diameter through the centre of the core, for the initial cycle (solid/dashed lines) and the second cycle (dotted lines). Column density spikes resulting from grain mantle re-injection at the discrete depth points have been smoothed over in the manner of GWR, assuming the ``maximal re-injection'' case. Hence, the column densities at these times represent their largest possible values.

The column density values and profile shapes of the first cycle are broadly repeated in the second, but with a few differences. Some species, most importantly carbon monoxide, clearly show enhancement due to the first cycle, at the beginning of the second. The other main difference is the diminution of hydrocarbon abundances at the ``shoulder'' occurring at around 0.72 Myr. The strength of this feature is dependent on the amount of carbon contained in CO. In the first cycle, at early times much carbon is present in the form of C$^+$, hence hydrocarbon (along with CO) abundances rise. This rise becomes steeper when depth points pass above the critical visual extinction. The resultant freeze-out of ionic molecules onto grain surfaces acts to lower the fractional ionisation and therefore lessen the effect of electron recombination on C$^+$ abundances. The main formation mechanisms of many hydrocarbons involve C$^+$ reactions, and the abundances of intermediary ions in the chain also benefit from the lower fractional ionisation, e.g. CH$_4$ below:
\\[0.17cm]
C$^+$, CH$^+$ $\stackrel{\mbox{\scriptsize{H$_2$}}}{\longrightarrow}$ CH$_{2}^{+}$ $\stackrel{\mbox{\scriptsize{H$_2$}}}{\longrightarrow}$ CH$_{3}^{+}$ $\stackrel{\mbox{\scriptsize{H$_2$}}}{\longrightarrow}$ CH$_{5}^{+}$ $\stackrel{\mbox{\scriptsize{CO,C}}}{\longrightarrow}$ CH$_4$
\\[0.17cm]
At times closer to peak density, freeze-out begins strongly to remove CO and hydrocarbons from the gas phase. CO is at this point the main source of carbon, and most C$^+$ is produced via reaction of CO with He$^+$; hence hydrocarbon production is diminished. The hydrocarbon downturn begins when the central depth point begins to suffer from very low free-carbon levels.

\begin{table}
\renewcommand{\thefootnote}{\fnsymbol{footnote}}
\centering
\begin{minipage}{55mm}
\caption{\label{tab2} Column density ratios between cycles 1 and 2, and cycles 1 and 3 (limit cycle), at $t=0.72$ Myr.}
\begin{tabular}{|l|c|c|}
\hline
 Species $i$ & N$_1$[$i$]:N$_2$[$i$] & N$_1$[$i$]:N$_3$[$i$] $^a$ \\
\hline
C$_3$H      &  9.2  &  10   \\
C$_5$H      &  17  &  18   \\
C$_3$H$_4$  &  45  &  55   \\
HC$_3$N     &  6.0  &  7.2   \\
SO          &  0.49  &  0.47  \\
SO$_2$      &  0.20  &  0.17  \\
\hline
\end{tabular}
\end{minipage}
\end{table}

The second cycle of the core does not show such a strong peak at the ``shoulder'' position, due to the retention of carbon in CO following the first cycle. Thus the chemical hysteresis resulting from the first cycle limits the chemistry of the second.

\begin{figure*}
\centering
\scalebox{0.9}[0.9]{\includegraphics{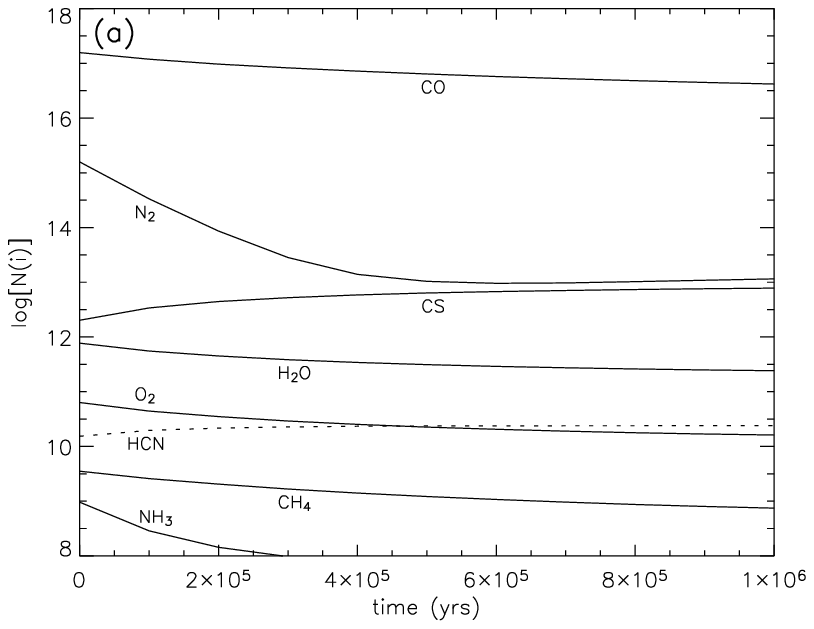}}
\scalebox{0.9}[0.9]{\includegraphics{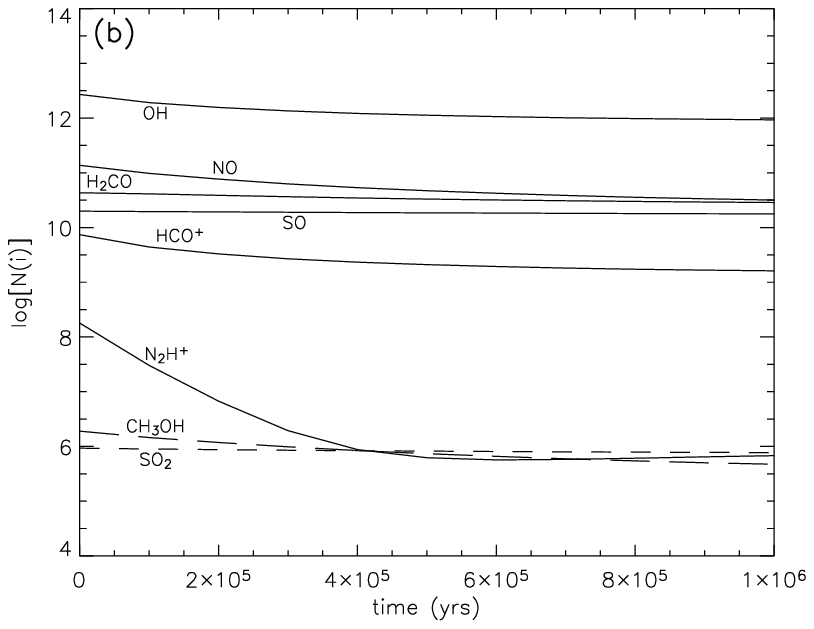}}
\caption{\label{fig2} Column densities as functions of time, calculated through the core, up to 1 Myr after first cycle run.}
\end{figure*}

Another difference present at the hydrocarbon ``shoulder'' period is the larger values of SO, and SO$_2$ (not shown, although see figure \ref{fig4}), in the second cycle. This is also due to the lower abundance of C/C$^+$ at these times and earlier (see figure \ref{fig1}b): SO is largely destroyed by reactions with atomic carbon:
\begin{eqnarray*}
\mbox{C + SO} & \rightarrow & \mbox{CS + O} \\
\mbox{C + SO} & \rightarrow & \mbox{CO + S}
\end{eqnarray*} 
These reactions, along with their C$^+$ analogues, account for between 50\% and 95\% 
of SO destruction at this time (dependent on depth). SO$_2$ is almost completely dependent on radiative association of SO with O, and neutral exchange with OH, for its formation. SO$_2$ is also largely destroyed by reaction with C$^+$.

The hydrocarbon species also exhibit weaker re-injection features (the ``hump'' beginning at approximately 1.2 Myr) in the second cycle, particularly at later times. The re-injection features end when all points in the core have fallen beneath the critical visual extinction. The latter parts of these features are caused by the re-injection of grain mantles from the innermost depth points. These are the points which retain most carbon in the form of CO after the first cycle, due to their higher visual extinctions. Because of this retention, there is less C/C$^+$ to freeze-out in the second cycle, and therefore much less CH$_4$ is formed on the grains. It is the re-injection of this CH$_4$ which leads to increased gas phase formation of most other hydrocarbons, and results in the ``re-injection features'' of those molecules. However, as mentioned previously, the uncertainty in the strength of these features makes them observationally unreliable.

Table \ref{tab1} shows column densities, and fractional abundances calculated from them, at the hydrocarbon ``shoulder'' ($t = 7.2 \times 10^{5}$ yrs) for the first, second and third cycles. Values which vary from the previous cycle by a factor of 2 or more are highlighted in bold face. The column densities of further cycles of the model are in fact very little different from the second, and almost all molecules show no difference after this cycle. By the third cycle, a limit cycle is quite definitely reached and almost all point by point fractional abundances (not shown) are identical at equal times through the cycle. Differences in column densities at peak time (1 Myr) are generally negligible, even for the second cycle, and hence are not shown. There are some significant differences between start and end time of the first cycle, as found by GWHRV. However, differences between later cycles are insignificant, hence we omit these results. We investigate the effects on these low density stages in more detail in the next section.

\subsection{Discussion}

We find that the effect of the successive cyclic processing of the gas has little obvious effect on the abundances in the dense gas of some of the most commonly observed molecules in dark clouds, i.e. CO, CS, HCO$^+$, N$_2$H$^+$ and NH$_3$. There are, however, significant effects on the abundances of some other molecules, at one fairly specific time in the evolution of a transient dense core.

Figure \ref{fig1} clearly shows that differences between the cycles are strongest at the ``shoulder'' feature, in particular for the hydrocarbons and SO/SO$_2$, and variations persist for longer (in terms of cycles) at this time. The time at which the ``shoulder'' feature peaks in various molecules is the same for each cycle, although individual molecules peak at slightly different times from each other; we choose $t=7.2 \times 10^{5}$ yrs as a representative time value. Table \ref{tab2} shows ratios between cycles for some strongly affected molecules. Hydrocarbons are suppressed by as much as 1 -- 2 orders of magnitude in later cycles; SO and SO$_2$ are enhanced by factors of approximately 2 and 5 respectively.

The differences in the ``shoulder'' feature may provide information on the frequency of the cyclic processing of the gas in observed dark clouds. High resolution observations which are capable of resolving individual cores (our basic units), or very small groups, in nearby dark clouds may reveal cores which happen to be at the hydrocarbon ``shoulder'' stage of their evolution, given a large enough sample of cores. We might estimate from figure \ref{fig1} that an unbiased sample of transient cores would yield some 5 -- 10 \% 
which were at that stage. Such cores would be characterised by high CO, CS and hydrocarbon levels, and would be of moderate visual extinction ($A_V \sim 2.5$ or more, from centre to edge) since the onset of freeze-out is important to the development of the ``shoulder'' feature. The abundances of hydrocarbon molecules, and their ratios with SO and/or SO$_2$ abundances could then be used to determine whether the chemistry were best described by a ``first cycle'' or ``later cycle'' model. 

We do not expect to be able to make any distinction between later cycles, within reasonable error margins. The distinction that could be made between ``first cycle'' and ``later cycle'' chemistries therefore corresponds to the cases of ``infrequent'' and ``frequent'' dynamical cycling of the gas, respectively. If cyclic processing of the gas occurs infrequently, then the chemistry of the gas has time to relax to an approximate steady state before any further processing is to take place. If cycling is frequent, then further processing takes place very soon, before such relaxation of the chemistry can occur, leading to the increased SO and SO$_2$ levels and lessened hydrocarbon abundances. The detection of ``first cycle'' (infrequent processing) chemistry could obviously also imply that core formation had not occurred at all before.

This dynamical history has only a very limited impact on the chemistry of the dense gas. However, if dense cores can be designated as ``first cycle'' or ``later cycle'' then we may take away information about the degree of chemical enhancement present in the diffuse background gas, as outlined above. The state of this gas would not be uniformly determinable by direct observation due to the low densities (and column densities) involved, and we should not expect to be able to detect cores at the very beginning or end of their cycles except in a few key molecules. (CN and NO both show large column density ratios between start and end time values in the first cycle -- 5.7 and 0.14 respectively -- but the absolute levels are on the margin of detectability.) 

\begin{figure*}
\centering
\scalebox{0.9}[0.9]{\includegraphics{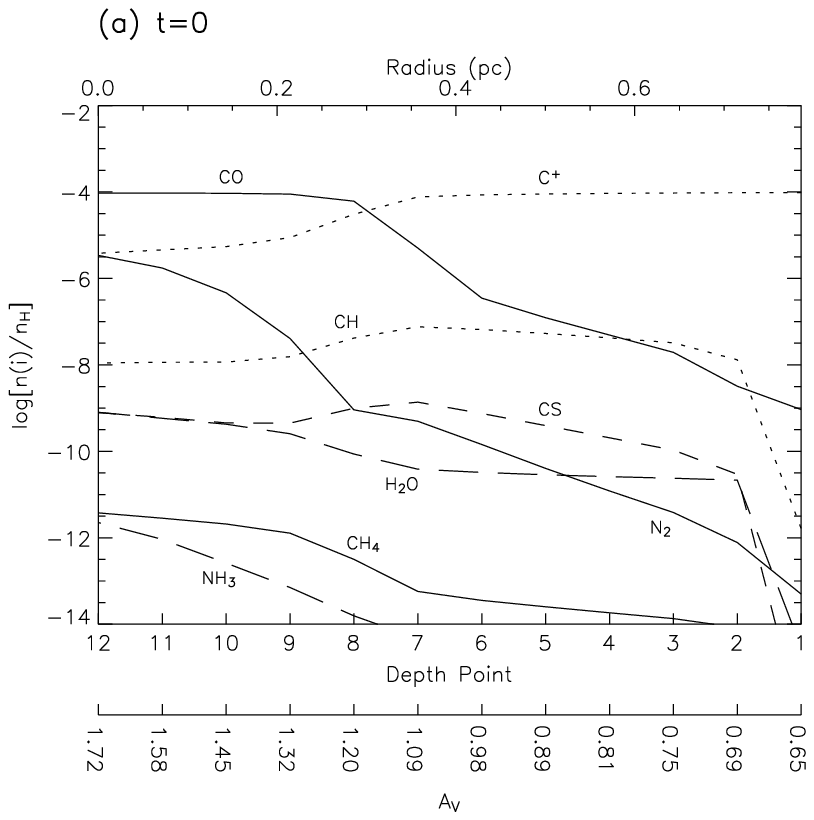}}
\scalebox{0.9}[0.9]{\includegraphics{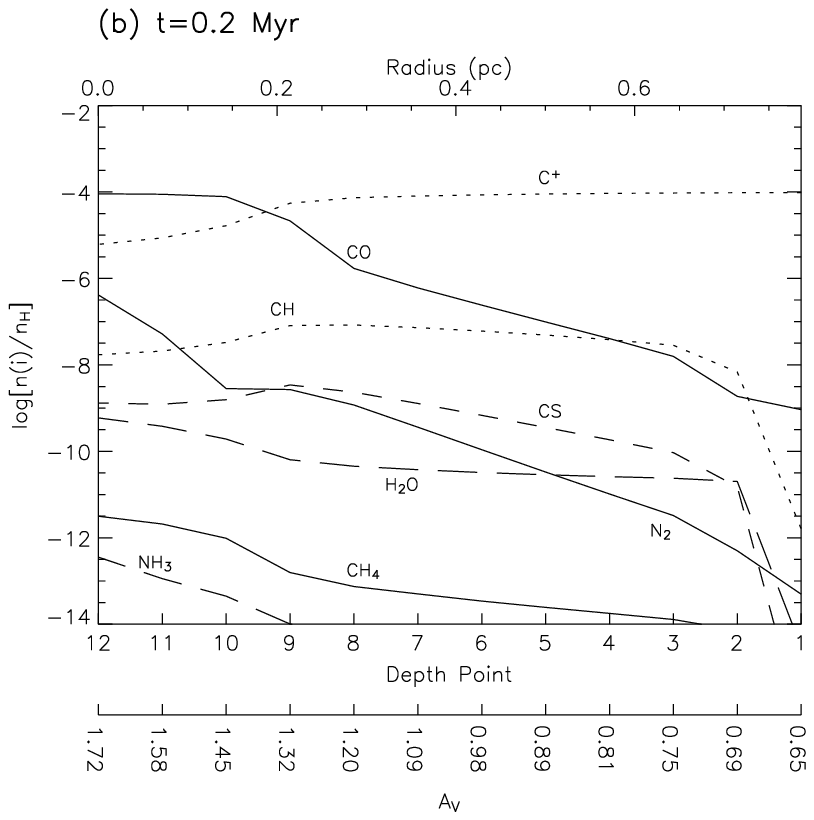}}
\scalebox{0.9}[0.9]{\includegraphics{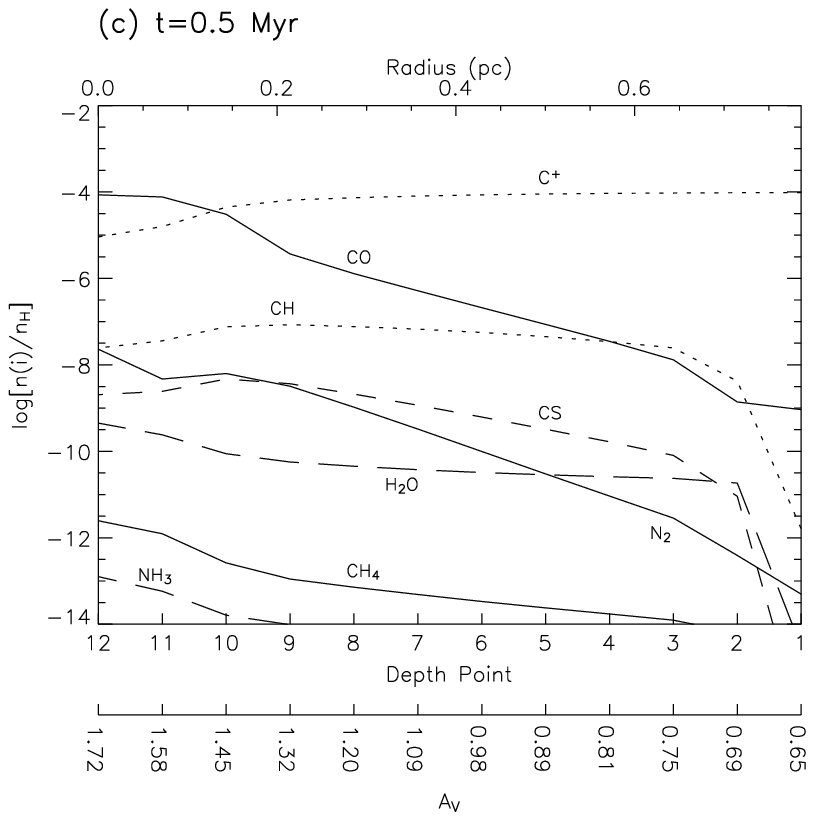}}
\scalebox{0.9}[0.9]{\includegraphics{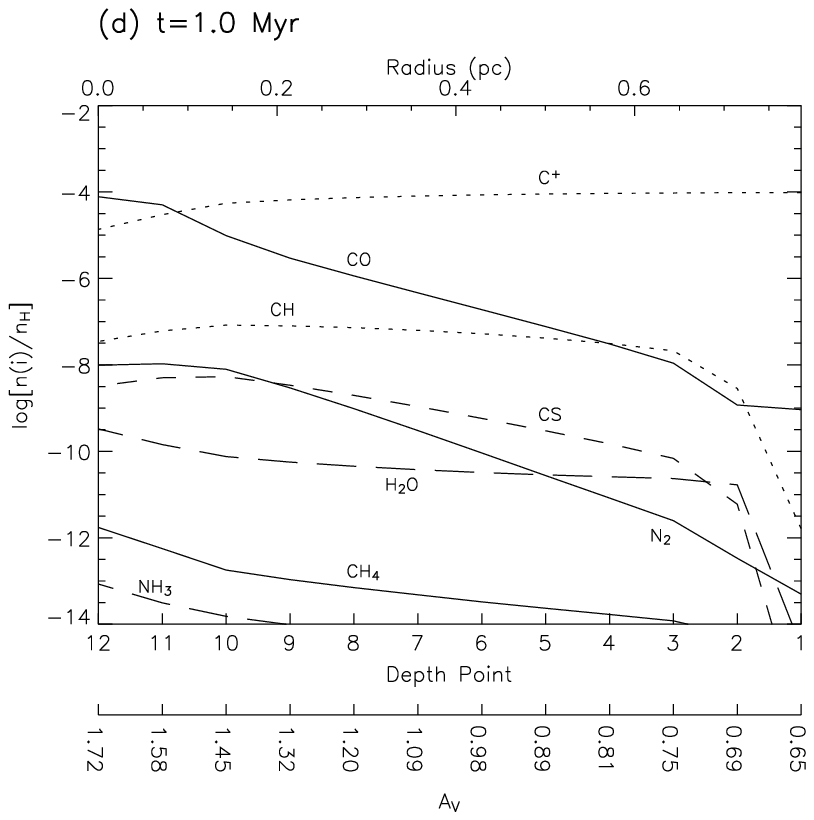}}
\caption{\label{fig3} Fractional abundances across the core -- up to 1 Myr after first cycle run (with no further dynamical evolution).}
\end{figure*}

\section{Processing of Diffuse Background Gas}

We now turn to the diffuse background gas from which the cores form, and into which they subsequently disperse. Does the chemical hysteresis have an effect on the diffuse gas, and how long will it endure?

We have already stated that a ``limit cycle'' is reached after 2 -- 3 cycles; indeed, comparing the end points of cycles, just one is required approximately to reach that state. Hence, continual cycling will not enhance the chemistry of the diffuse gas any further. We may, however, still distinguish between that material which has been recently processed and that which has not. The solid lines in figure \ref{fig1} illustrate the difference between the unprocessed ($t=0$) and the processed diffuse gas ($t=2$ Myr). How long can we expect these enhancements to last, and how spatially extended will they be? And therefore, in a regime where cycling of the gas is taking place, how frequently must the gas be processed and re-processed for it to remain constantly enhanced over equilibrium levels?

GWHRV gave consideration to the effects on the diffuse gas, however those results went only as far as the end of one cycle, and did not consider the long-term chemical enhancement of the gas.

In order to answer the above questions, we take the output values of fractional abundances from all depth points at the end time of the first cycle as a starting point for a further simulation. Using these values, we maintain the physical state of the core at the end of the first cycle run (i.e. diffuse conditions, see GWHRV) and run the chemical code until a steady state has been achieved. No freeze-out onto dust grains takes place, as visual extinctions at all points are below the critical value, $A_{V,crit}=2.5$.

\subsection{Results}

Figure \ref{fig2} shows column densities through the core gas up to 1 Myr after the core's physical evolution has come to a halt. 

By a time of approximately 0.2 Myr into this run, those species which show non-steady state values after the first cycle run are still more than an order of magnitude different from their pre-cycle steady-state values. After half a million years, values are close to steady state.

The longevity of molecular material this long after the end of the cycle is a result of the multi-point approach of the model, and the density and visual extinction structure adopted (see equations 1 \& 2, and figure \ref{fig0}). Figures \ref{fig3}a -- d show point by point fractional abundances at four times in this post-first cycle run. These plots show that the calculated column densities are strongly dependent on the penetration of the interstellar radiation field into the post-cycle core. As time passes, the chemistries at individual depth points are affected at different rates according to the static visual extinction profile. Whilst the inner depth points have higher visual extinctions which protect molecular species from photodestruction, the higher densities at these points expedite ion-molecule reactions which are fed by photoionisation. Hence the combination of these two static profiles governs the gradient of molecular degeneration across the core.

Figure \ref{fig3}a shows the extent of molecular material at the end time of the first cycle run. Figure \ref{fig3}b shows its extent after another 200,000 years. At this point, although the extent of CO has fallen by approximately half, from 0.3 pc to 0.15, this extent represents an unattenuated fractional abundance of about 10$^{-4}$. Material with $X($CO$) \geq 10^{-5}$ still extends out to $\sim$0.25 pc at this time, and to $\sim$0.20 pc at 0.5 Myrs, compared to $\sim$0.35 pc at the end of the first cycle run. This effect is more pronounced for CO, due to its self-shielding at visual extinctions greater than about 1, however the same effect is seen in all of the enhanced species; long-lived abundances significantly higher than the usual diffuse levels, extending far into the surrounding dark cloud. Table \ref{tab2b} shows the extents of molecules at various times for typical enhanced fractional abundance values. Besides CO, O$_2$ and H$_2$O are particularly extended in comparison with the start of the first cycle (i.e. steady state levels).


\begin{table}
\renewcommand{\thefootnote}{\fnsymbol{footnote}}
\centering
\begin{minipage}{80mm}
\caption{\label{tab2b} Typical enhanced fractional abundances with respect to $n_{\mbox{\scriptsize{H}}}$, and their extents.}
\begin{tabular}{|l|c|c|c|c|}
\hline
 Species $i$ & log$[X(i)]$ & & Extent (pc) & \\
   &  &  \multicolumn{2}{c}{1st cycle} & After 1st cycle \\
   &  & (Start) & (End) & ($t=0.5$ Myr) \\
\hline
CO       &  -5   &  0.13   &  0.35  &  0.19   \\
CH$_4$   &  -12  &   ---   &  0.23  &  0.084  \\
H$_2$CO  &  -11  &  0.16   &  0.27  &  0.18   \\
H$_2$O   &  -10  &  0.089  &  0.28  &  0.14   \\
O$_2$    &  -11  &  0.046  &  0.24  &  0.11   \\
NO       &  -11  &  0.13  &  0.28  &  0.15   \\
\hline
\end{tabular}
\end{minipage}
\end{table}

\begin{figure}
\centering
\scalebox{0.9}[0.9]{\includegraphics{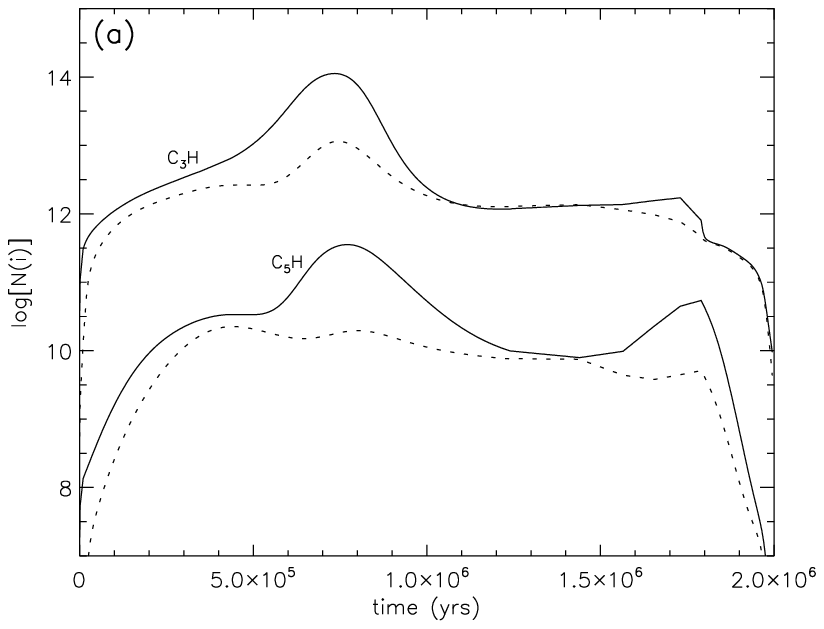}}
\scalebox{0.9}[0.9]{\includegraphics{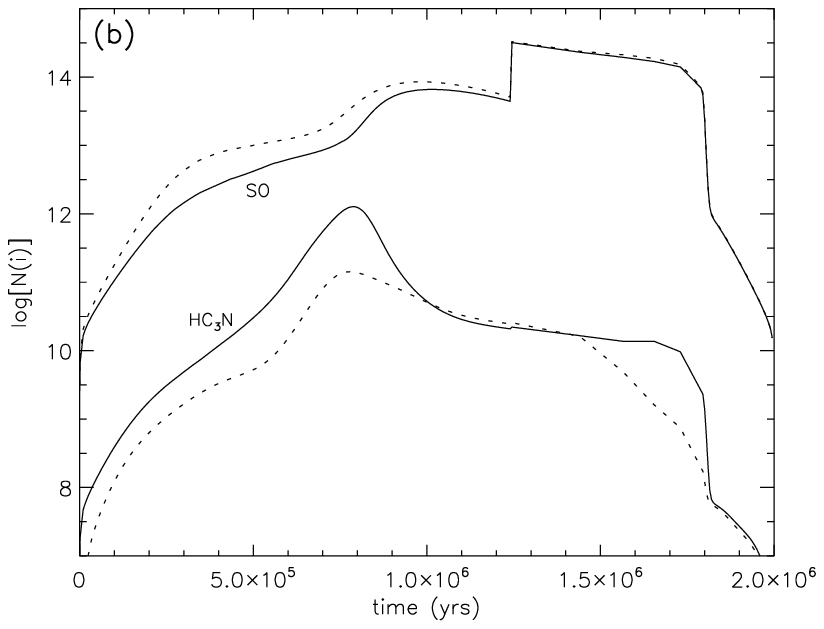}}
\scalebox{0.9}[0.9]{\includegraphics{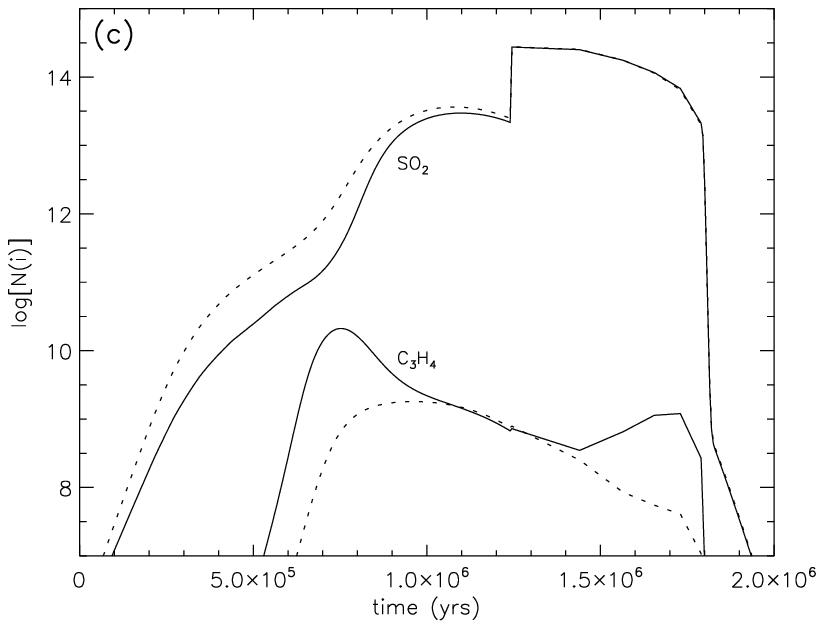}}
\caption{\label{fig4} Column densities of selected transitions as functions of time -- first cycle chemistry (solid line) and limit cycle chemistry (dotted line)}
\end{figure}

\subsection{Discussion}

The longevity of enhancements in column density values (figure \ref{fig2}) is strongly dependent on the levels at individual depth points in the innermost parts of the cores, which retain significant amounts of molecular material formed at high density/$A_V$ times. The outer points generally lose most of their molecular material before the end of a cycle is reached, so levels at these points do not affect the behaviour of any subsequent core cycles. The reduction in the extent of molecular material is controlled by local radiation field and density conditions.

\begin{figure}
\centering
\scalebox{0.50}[0.50]{\includegraphics{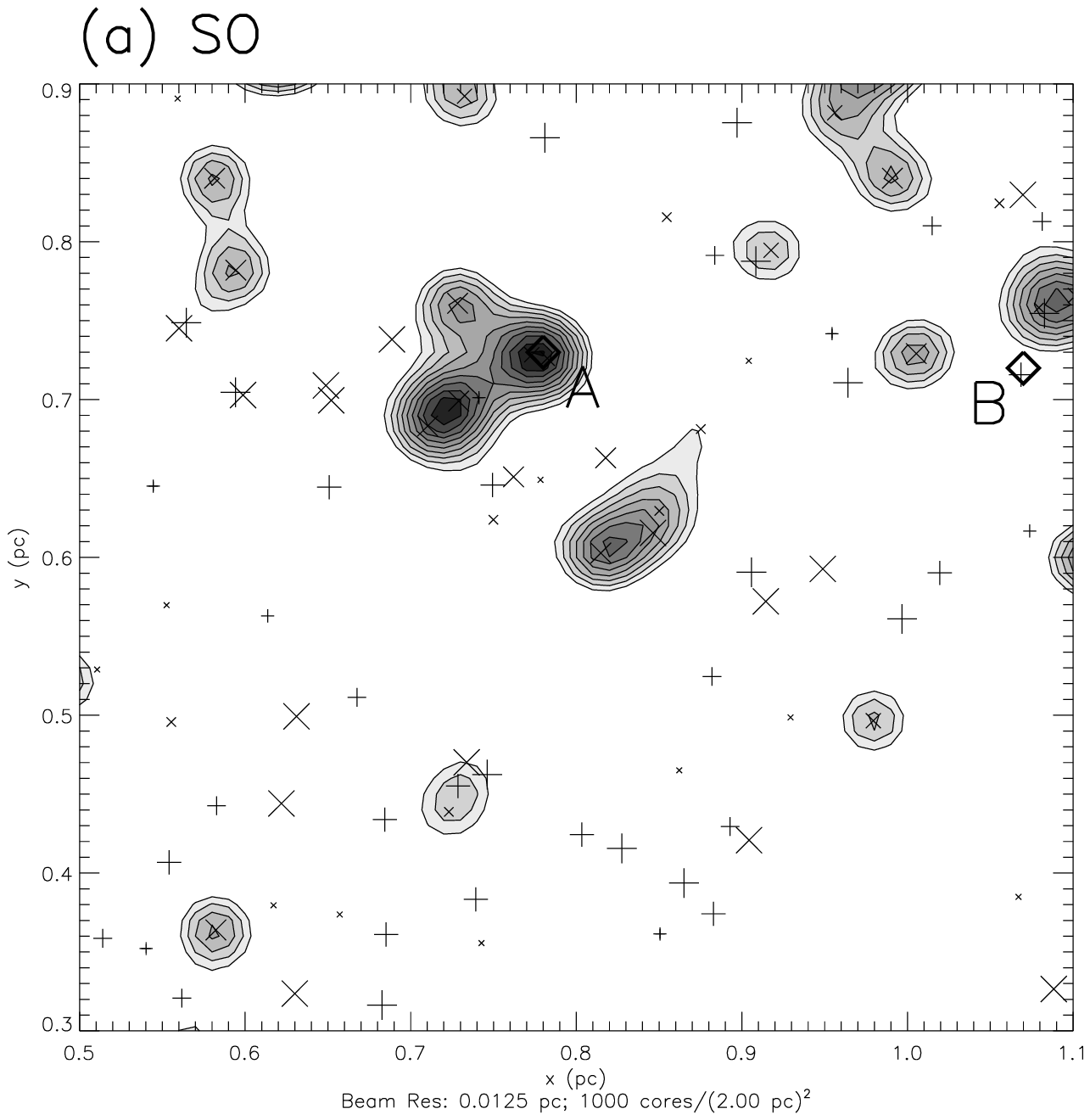}}
\scalebox{0.50}[0.50]{\includegraphics{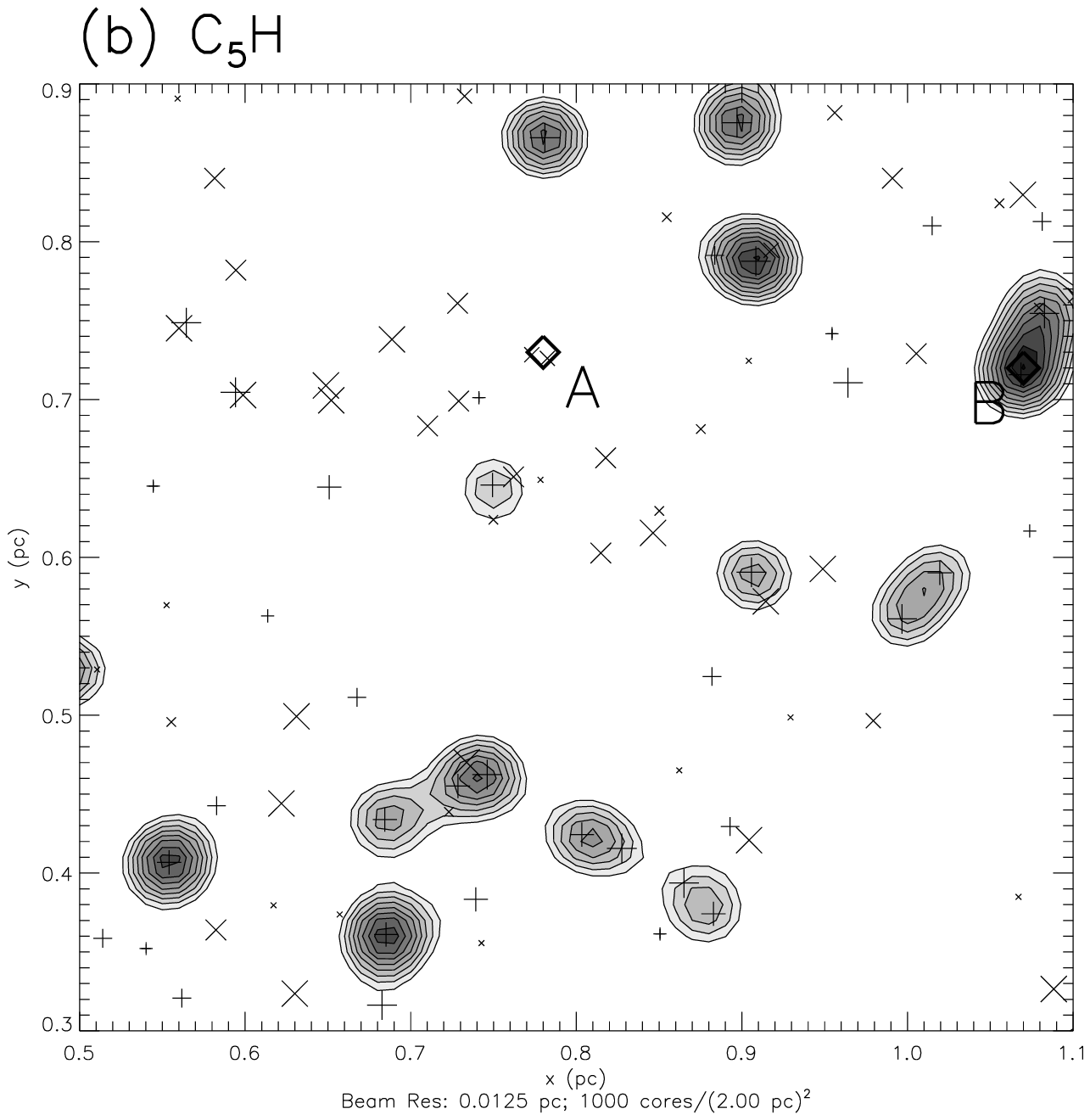}}
\caption{\label{fig5} High resolution convolved column density maps for cores with first cycle chemistries.}
\end{figure}

\begin{figure}
\centering
\scalebox{0.50}[0.50]{\includegraphics{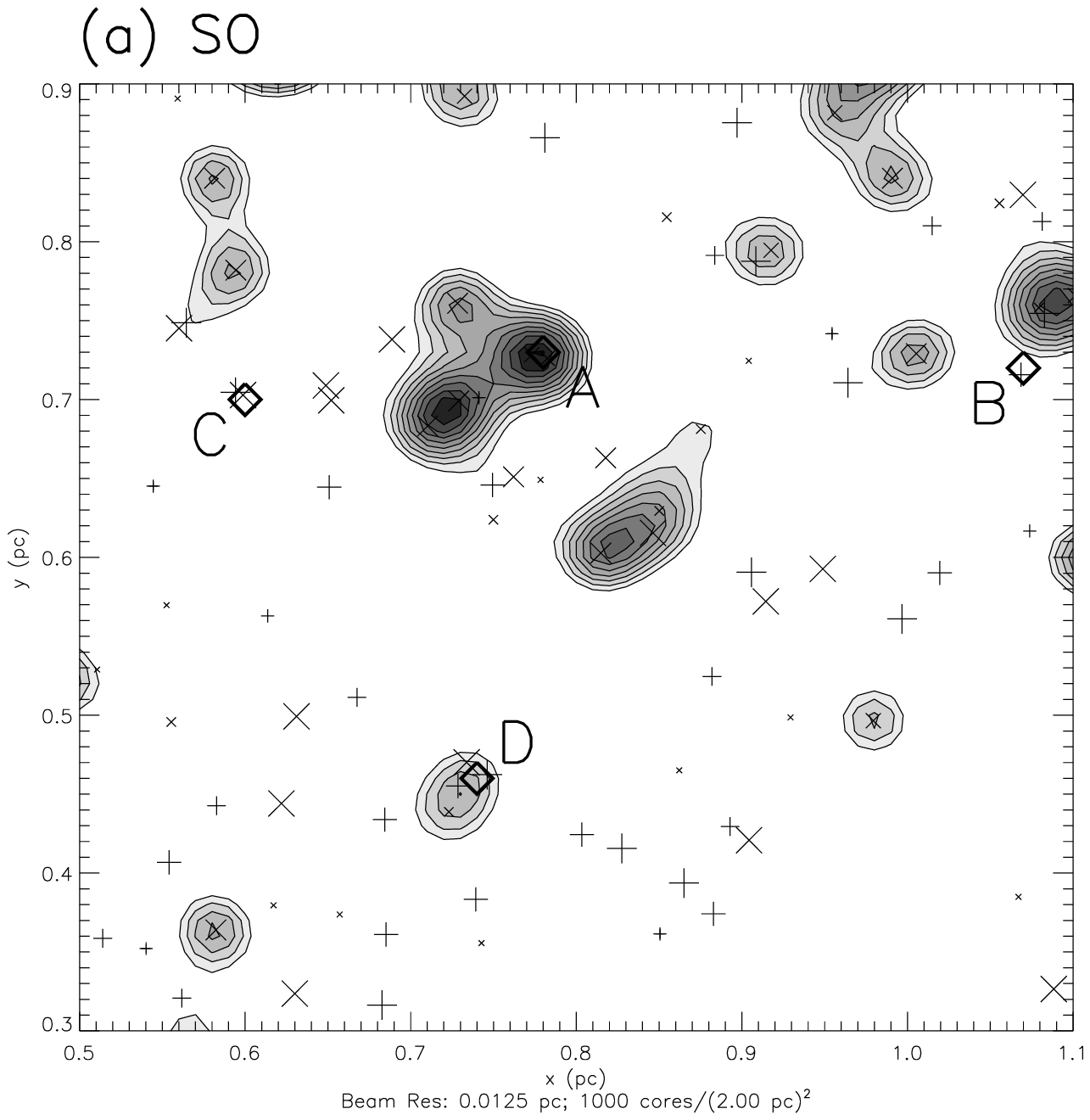}}
\scalebox{0.50}[0.50]{\includegraphics{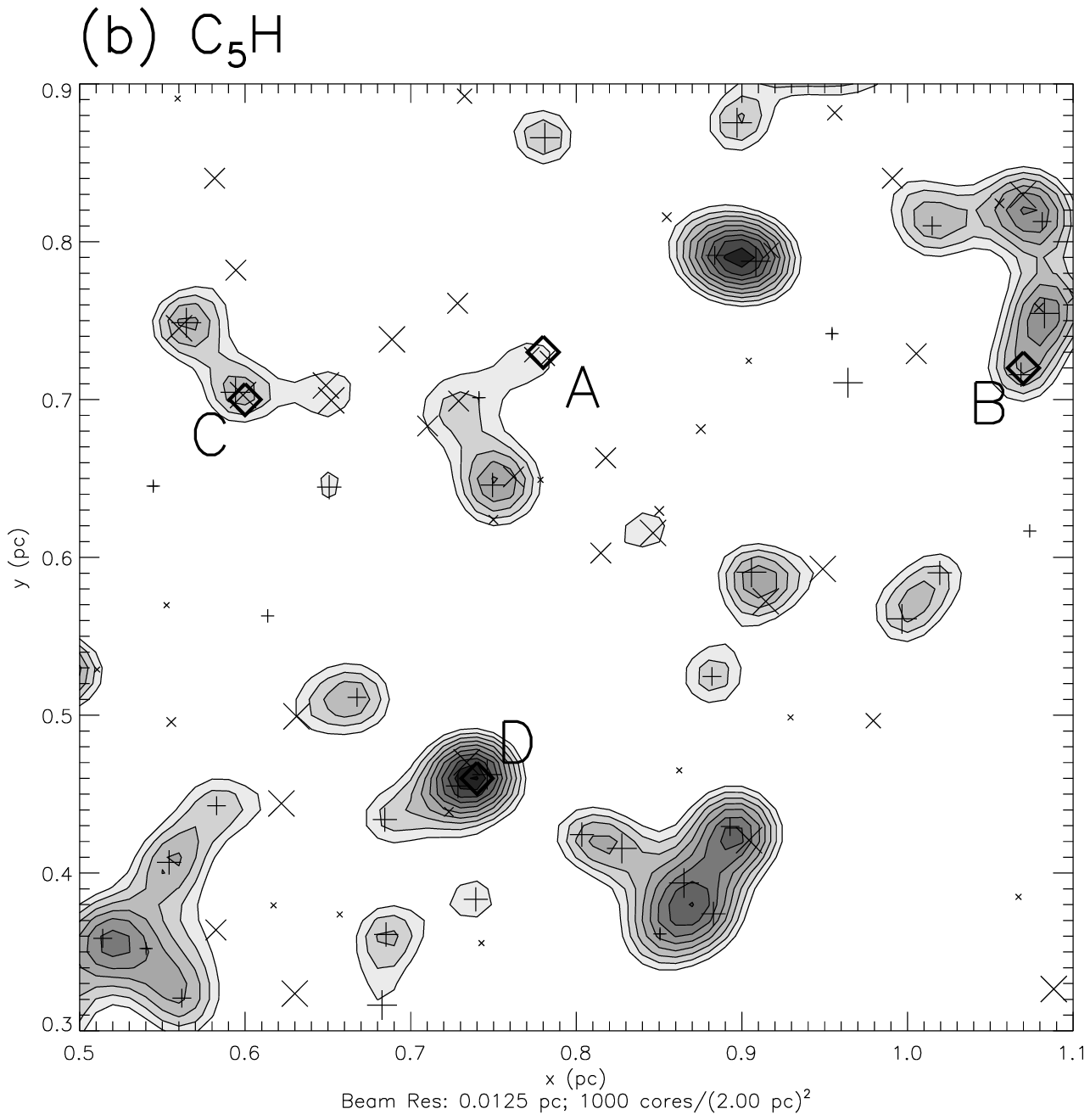}}
\caption{\label{fig6} High resolution convolved column density maps for cores with limit cycle chemistries.}
\end{figure}

The attainment of either first cycle or later cycle chemistry would be dependent on the period of time between the end of one cycle and the onset of another. We may attribute a maximum period of $\sim$0.5 Myrs between cycles for the subsequent chemistry to show appreciable signs of previous enhancement. This is the point (after end time) at which the innermost depth points begin to return to their equilibrium values -- these points contribute the most to column densities at the hydrocarbon ``shoulder'', and so when their carbon monoxide is broken down they may produce more hydrocarbons from the free carbon, in their subsequent cycle. 

We find, like GWHRV, that material formed at high densities is carried out to large distances. However, the longevity of the molecular material indicates that this process could be very important. This gives core formation and destruction much more than just a local effect; subsequent core formation (or other processes) may further move, cycle and/or chemically process the already enhanced molecular gas.

On this basis, evidence of small-scale structure in clouds could indicate a non-equilibrium chemistry in the diffuse background gas, especially if the high-density signatures of multiple-cycle processing can be observed. Detection of later cycle chemistries would inherently imply that the intercore background gas should be strongly chemically enhanced. 

Alternatively, widespread detections of molecular enhancement (of the sort shown in table \ref{tab2b}) in translucent clouds or in the translucent envelopes of giant molecular clouds would also indicate a ``frequent'' cycling process. Indeed, such cyclic core formation and chemical processing may help to explain the variability in the ratio of CO line strengths to molecular hydrogen column densities in translucent clouds, as evidenced by e.g. Magnani et al. \nocite{magnani03a}(2003, and references therein).

\section{Effects of Core Cycling on Molecular Line Maps}

The treatment by GWR of a molecular cloud as being composed of an ensemble of dense cores produced compelling agreement with observational results. GWR applied column density information obtained from a single core model to an ensemble of randomly positioned cores, convolving the resultant column density map to produce synthetic molecular emission maps at both high and low angular resolutions. These maps agreed well with general observed morphologies of dark clouds, and produced plausible column densities for CO, CS, HCO$^+$, N$_{2}$H$^{+}$ and NH$_3$. 

Using this synthetic mapping technique, we now put to the test the observational effects which we have predicted in section 2. Are there observational tracers to distinguish between ``first cycle'' and ``later cycle'' cores?

To do this we apply the randomly generated basic ``sky'' map used by GWR, to two limiting cases. In the first, we attribute to each core in the map a ``first cycle'' chemistry. Each core has associated with it a randomly generated age, anywhere between the start and end time of the chemical model (0 -- 2 Myr). We use this age with the results from section 2 to determine the column densities associated with each core. In the second case, we use instead the ``limit cycle'' chemistry obtained from the third dynamical cycle of section 2; everything else, including the cores' ages, remains the same. In each case we convolve the column densities using a high resolution beam-width, producing molecular ``emission'' maps. Simulated FWHM beamwidth is commensurate with that of the high resolution observations of \cite{morata03a}: $\sim$20 arcsec at a distance of $\sim$300 pc, as adopted in GWR. The mapped region corresponds to that selected by GWR for the ``high resolution'' study, chosen because of strong features in CO, CS, HCO$^+$, N$_2$H$^+$ and NH$_3$.

We map the hydrocarbon species C$_3$H$_4$, C$_5$H, and HC$_3$N, along with SO, being among the most strongly differentiated species at the hydrocarbon ``shoulder''. It would be desirable to map H$_2$CO, however the effective critical density provided by \cite{evans99a} of $6 \times 10^{4}$ cm$^{-3}$ is too high for the densities our chemical model employs ($5 \times 10^{4}$ cm$^{-3}$) to produce detectable emission (deemed to be a 1K line). CH$_4$ is also strongly affected at the ``shoulder'' feature, but has no dipole moment and hence is very difficult to observe. We assume effective critical density values of $n_{\mbox{\scriptsize{{\em eff}}}}=10^{3}$ cm$^{-3}$ for the four mapped species as for CO (see GWR), in the absence of other information. ``Detected'' column densities therefore represent ``emission'' from core gas of $n_H > n_{\mbox{\scriptsize{{\em eff}}}}$.

\begin{table}
\renewcommand{\thefootnote}{\fnsymbol{footnote}}
\centering
\begin{minipage}{58mm}
\caption{\label{tab3} Peak column densities for first cycle maps.}
\begin{tabular}{|l|c|c|c|c|}
\hline
 Molecule & x & y & $N$[$i$] $^{a}$ & Peak $^{b}$ \\
 & (pc) & (pc) & (cm$^{-2}$) &  \\
\hline
SO          &  {\bfseries 0.78}  &  {\bfseries 0.73}  &  {\bfseries 3.5(14)} & {\bfseries A} \\
            &  1.07  &  0.72  &  6.1(13) & B \\
\hline
HC$_3$N     &  0.78  &  0.73  &  2.5(10) & A \\
            &  {\bfseries 1.07}  &  {\bfseries 0.72}  &  {\bfseries 9.3(11)} & {\bfseries B} \\
\hline
C$_3$H$_4$  &  0.78  &  0.73  &  5.6(08) & A \\
            &  {\bfseries 1.07}  &  {\bfseries 0.72}  &  {\bfseries 1.6(10)} & {\bfseries B} \\
\hline
C$_5$H      &  0.78  &  0.73  &  1.5(10) & A \\
            &  {\bfseries 1.07}  &  {\bfseries 0.72}  &  {\bfseries 2.9(11)} & {\bfseries B} \\
\hline
\end{tabular}
$^{a}$$A(B)=A \times 10^{B}$ \\
$^{b}$Peaks in bold face correspond to the peak in the given molecule. \\
\end{minipage}
\end{table}

Figure \ref{fig4} shows the time-dependent column densities of some observable species which show strong enhancement at the hydrocarbon ``shoulder'' feature, as calculated using the models of section 2. Re-injection features are smoothed to take maximal values (see GWR). SO$_2$ shares broadly similar column density features to SO, and their resultant maps are very similar, hence SO$_2$ is omitted from the mapping procedure. SO also exhibits slightly stronger column density enhancements, as calculated from convolved values. C$_5$H exhibits a starker contrast between the two cases than the smaller molecule C$_3$H, since it is dependent on two more C$^+$ reactions for its formation at the ``shoulder'' time (see section 2.1). Hence it provides better morphological distinction between the two cases.

The ``shoulder'' feature is generally the overall hydrocarbon column density peak in the first cycle case, so map peaks in these molecules should correspond to that time. As stated in section 2.2 we should therefore expect $\sim$5 -- 10 \% 
of cores to show emission in these molecules. For the limit cycle case, the hydrocarbons tend to peak at other times, so should not necessarily map the same regions.

\subsection{Results}

Figures \ref{fig5} and \ref{fig6} show convolved maps of SO and C$_5$H, for the first cycle and limit cycle cases, respectively. (HC$_3$N and C$_3$H$_4$ were also mapped, but show little morphological differentiation between the first cycle and limit cycle cases.) 

Tables \ref{tab3} \& \ref{tab4} show convolved column densities determined at the peaks for each molecule (bold face indicates the peak corresponding to the molecule), for first cycle and limit cycle maps, respectively. The hydrocarbon molecules all share the same peak position (B) for the first cycle case, indicating their ``shoulder'' feature origin. The strongest SO peak is at an entirely different position (A). The SO morphology closely resembles that of NH$_3$ shown in GWR (their fig. 5e). The sharpness of the column density peaks shown in figure \ref{fig4} produces very sharp convolved contours for each molecule. This is because only a small number from the sample of cores is at a stage of evolution corresponding to the molecule's column density peak, therefore few cores are in close enough proximity at this high angular resolution to be convolved together.

\begin{table}
\renewcommand{\thefootnote}{\fnsymbol{footnote}}
\centering
\begin{minipage}{58mm}
\caption{\label{tab4} Peak column densities for limit cycle maps.}
\begin{tabular}{|l|c|c|c|c|}
\hline
 Molecule & x & y & $N$[$i$] $^{a}$ & Peak $^{b}$ \\
 & (pc) & (pc) & (cm$^{-2}$) &  \\
\hline
SO          &  {\bfseries 0.78}  &  {\bfseries 0.73}  &  {\bfseries 3.6(14)} & {\bfseries A} \\
            &  1.07  &  0.72  &  8.2(13) & B \\
            &  0.60  &  0.70  &  1.2(14) & C \\
            &  0.74  &  0.46  &  1.8(14) & D \\
\hline
HC$_3$N     &  0.78  &  0.73  &  2.3(10) & A \\
            &  {\bfseries 1.07}  &  {\bfseries 0.72}  &  {\bfseries 1.2(11)} & {\bfseries B} \\
            &  0.60  &  0.70  &  7.0(10) & C \\
            &  0.74  &  0.46  &  1.1(11) & D \\
\hline
C$_3$H$_4$  &  0.78  &  0.73  &  3.8(08) & A \\
            &  1.07  &  0.72  &  8.3(08) & B \\
            &  {\bfseries 0.60}  &  {\bfseries 0.70}  &  {\bfseries 2.5(09)} & {\bfseries C} \\
            &  0.74  &  0.46  &  2.1(09) & D \\
\hline
C$_5$H      &  0.78  &  0.73  &  1.2(10) & A \\
            &  1.07  &  0.72  &  1.7(10) & B \\
            &  0.60  &  0.70  &  1.6(10) & C \\
            &  {\bfseries 0.74}  &  {\bfseries 0.46}  &  {\bfseries 3.0(10)} & {\bfseries D} \\
\hline
\end{tabular}
$^{a}$$A(B)=A \times 10^{B}$ \\
$^{b}$Peaks in bold face correspond to the peak in the given molecule. \\
\end{minipage}
\end{table}

\begin{table*}
\renewcommand{\thefootnote}{\fnsymbol{footnote}}
\centering
\begin{minipage}{117mm}
\caption{\label{tab5} Molecular ratios at various column density peaks, for the standard case ($R_{stan}$) and the limit cycle case ($R_{lim}$).}
\begin{tabular}{|l|c|c|c|c|c|}
\hline
 Molecular Ratio & $R_{stan}$ $^{a,b}$ & $R_{lim}$ $^{a,b}$ & $R_{stan}$/$R_{lim}$ & Expected $R_{stan}$/$R_{lim}$ $^{c}$ & Peak $^{d}$ \\
\hline
{\bfseries HC$_3$N}:SO     &  {\bfseries 1.53(-2)}  &  {\bfseries 1.41(-3)}  &  {\bfseries 10.90}  &  {\bfseries 15.5}  & {\bfseries B} \\
                           &             7.31(-5)   &             6.49(-5)   &              1.13   &    &            A \\
\hline
{\bfseries C$_3$H$_4$}:SO  &  {\bfseries 2.61(-4)}  &  {\bfseries 2.07(-5)}  &  {\bfseries 12.62}  &  {\bfseries 117.3}  & {\bfseries B, C} \\
                           &             1.62(-6)   &             1.06(-6)   &              1.54   &    &            A \\
\hline
{\bfseries C$_5$H}:SO      &  {\bfseries 4.77(-3)}  &  {\bfseries 1.61(-4)}  &  {\bfseries 29.58}  &  {\bfseries 38.0}  & {\bfseries B, D} \\
                           &             4.19(-5)   &             3.26(-5)   &              1.29   &    &            A \\
\hline
\end{tabular}
$^{a}$$A(B)=A \times 10^{B}$ \\
$^{b}$$R = N[i]/N[j]$ \\
$^{c}$ Calculated from hydrocarbon ``shoulder'' values shown in table 2. \\
$^{d}$ Peaks in bold face correspond to molecule in bold face. Dual peak listings indicate the peak for the first cycle (standard) run and the peak for the limit cycle run, respectively. \\
\end{minipage}
\end{table*}

The mapping of limit cycle cores in SO results in few morphological differences from the first cycle case, and the peak column density is almost identical (see tables \ref{tab3} \& \ref{tab4}). SO column density is a little different comparing the two cases at peak B, the HC$_3$N peak. Morphologically, HC$_3$N and C$_5$H are similar, and in the limit cycle case, they map some of the same regions as they do in the first cycle case (associated with CS detection), but also some of the central ``ridge'' regions, marked most strongly by N$_2$H$^+$ and HCO$^+$ in GWR. This behaviour is similar to that of CO, and indicates the stronger weighting towards middle times in the column density profiles of figure \ref{fig4}.

Peaks B, C, and D for the limit cycle case correspond to the peaks in HC$_3$N, C$_3$H$_4$, and C$_5$H, respectively. C$_5$H shows the greatest fall in convolved peak column density (comparing limit cycle peak D with first cycle peak B). Meanwhile, SO shows its greatest increase at peak D. Hence the C$_5$H:SO ratio is very much greater for the first cycle C$_5$H peak (B) than for the limit cycle C$_5$H peak (D) -- a factor of $30 \times$. Other such molecular ratio information is shown in table \ref{tab5}. Other hydrocarbon:SO ratios are around an order of magnitude greater at hydrocarbon peaks for the first cycle case than the limit cycle case.

\subsection{Discussion}

We find several observational markers for the distinction of the first cycle case and the limit cycle case. Morphological comparison with the results of GWR suggest that C$_5$H peak positions and map coverage which match closely with CS transitions would indicate a first cycle chemistry, whereas closer agreement with the more widespread CO coverage would indicate the limit cycle case. The morphologies of C$_3$H$_4$ and HC$_3$N should behave similarly, however the sparsity of coverage may make their behaviour difficult to discern.

C$_5$H should also be the key molecule to compare with SO abundances. Table \ref{tab5} shows the molecular ratios derived from column densities at particular peaks, for the standard case ($R_{stan}$) and for the limit cycle case ($R_{lim}$). C$_5$H:SO ratios at C$_5$H peaks should allow the distinction of first cycle and limit cycle chemistries. Even if the total C:S ratio is in doubt \cite[as could be the case -- hydrocarbon abundances in this model are generally somewhat lower than are observed in TMC-1 core D; see][]{smith04a}, the C$_5$H:SO ratios at the SO peaks should provide baselines. These ratios do not vary substantially between the first cycle and limit cycle cases.

At lower resolutions than those used here, the morphological comparisons still hold (i.e. comparative coverage of hydrocarbons as compared to CO, CS, etc.) however, the contrast in molecular ratios (see table \ref{tab5}) becomes progressively more ``washed out'' in comparison with the expected (unconvolved) ratios obtained from the models. This is because as lower angular resolutions result in the convolution of more cores, averaging out their contributions. Table \ref{tab5} shows the ratios (hydrocarbon:SO)$_{stan}$ / (hydrocarbon:SO)$_{lim}$ expected using the hydrocarbon ``shoulder'' ratios shown in table 2. The convolved ratio for C$_5$H is not much lower than the expected value. However, for C$_{3}$H$_4$ the ratio is much reduced from that predicted in table \ref{tab5}. This is because in the limit cycle case the ``shoulder'' is so diminished that it is no longer the dominant contributor to detected column densities.

For convolutions using the low resolution beam simulated in GWR (FWHM $\simeq$ 2 arcmin, not shown), the ratio is further reduced to (C$_{5}$H:SO)$_{stan}$ / (C$_{5}$H:SO)$_{lim}=12.5$ at the C$_{5}$H peak, whilst at the SO peak it reaches a substantial $4.2$. Hence, the observable difference between first cycle and limit cycle cases is diminished, and the baseline by which it is judged is less reliable (assuming C:S ratios are uncertain).

As stated in section 3, such a determination of limit cycle chemistry over the first cycle chemistry would imply a state of chemical enhancement in the diffuse background gas. This in turn would limit the time period between the beginning of one cycle and the onset of the next to $<$$0.5$ Myr. Such a situation would indicate that the initial chemical conditions commonly adopted in dark cloud models would be far too simplistic and more open to local variability.

We have tested two limiting cases, wherein all cores exhibit either a first cycle or a limit cycle chemistry. However, we cannot rule out that a single cloud could contain different regions where either the first cycle or limit cycle case is typical.

In these models we have necessarily had to limit our investigation to one set of transient core parameters. Most crucial to the effects we have investigated here is the ability to form significant quantities of CO during the dense stage, and to retain it for substantial periods in the central parts of the post-dissipation core gas. Such formation clearly must occur, due to its wide-spread detection in dense clouds. Its survival in the gas phase after core dissipation requires a visual extinction of around 1 or more. Hence, whilst the parameter space of this model is potentially large, the broad behaviour investigated here should be quite ubiquitous. Test runs have also shown the model to be quite robust to changes in the dynamical timescale of at least a few times longer or shorter. We have not yet investigated the (quite probable) situation where a cloud contains a spectrum of (maximum) core sizes and masses. Further improvements in our understanding of the dynamical processes involved will allow the parameter space to be narrowed.

\section{Conclusions}

The conclusions of previous papers on core cycling in dark clouds are preserved in this work; the chemistry of cores is characteristically young (and remains so in subsequent cycles), and the re-injection of grain mantle-bound molecular material strongly affects abundances. 

In addition, we draw the following main conclusions from this study:

\begin{enumerate}

\item A limit cycle is reached after 2 -- 3 cycles of the GWHRV process, achievable well within dark cloud lifetimes.

\item Little chemical difference is found between first cycle and limit cycle chemistries in the dense gas, save at the pre-peak time (ie. pre-maximum density) hydrocarbon ``shoulder''. Peak time chemistries of different cycles cannot be distinguished. 

\item We suggest observational indicators to determine whether cores are in their first dynamical cycle or a later cycle, using ratios of hydrocarbons to SO or SO$_2$, with the higher values signifying a first cycle chemistry. In particular, C$_5$H:SO ratios should be sought. Observed morphologies should also vary somewhat.

\item Only high resolution surveys capable of resolving individual transient cores should be capable of distinguishing between first cycle and later cycle chemistries in dark clouds. Only a fraction of observed transient cores are expected to be at an evolutionary stage at which their chemistries may indicate their dynamical histories. Chemical signatures would likely be lost at lower beam resolutions where small-scale structure was not well resolved.

\item Enhanced molecular abundances in the diffuse background gas endure for long periods ($\sim$0.5 Myr) subsequent to core formation/dissipation, and extend to large distances ($\sim$$2 \times$ equilibrium extents, or more). 

\item In a scenario where the gas is never dynamically dormant for more than $\sim$0.5 Myr before cyclic processing ensues, a non-equilibrium chemistry should be ubiquitous in the diffuse inter-core regions from which the denser regions form. Column densities in the inter-core regions may be too low to distinguish such a scenario, but positive detection of the observational indicators proposed above, in the dense gas, would be compelling. If intercore regions are detectable as such, then a higher NO:CN ratio could indicate an enhanced, post-core state for the gas. 

\item If only first cycle chemistry is detected in transient cores in dark clouds then we should expect little enhancement to the diffuse background gas, or only localised enhancement.

\item The existence of such non-equilibrium chemistry in the diffuse background may have consequences for the initial conditions commonly used in chemical models of interstellar clouds.

\end{enumerate}

The observations of L673 by \cite{morata03a} could indicate such a case of background chemical enhancement in a dark cloud. The separations between cores in that study are generally less than 1 arcmin for the CS, and those seen in HCO$^+$ and N$_2$H$^+$ show similar separations, corresponding to inter-core distances potentially as low as 0.09 pc. Line centre measurements in that study were compatible with the cores being spatially associated. A multi-transition study (Morata, Girart \& Garrod, in prep.) has recently been carried out for region L673 at points corresponding to both clump and inter-clump material. Molecules surveyed include SO, SO$_2$, C$_2$H and C$_3$H$_2$, as well as CN and NO. We hope that this study may provide observational evidence in support of the ideas put forward here.

Observations \nocite{smith04a}(see Smith et al. 2004 for a collection of previous observational results) and chemical models \cite[e.g.][]{hartquist01a} of TMC-1 core D indicate both a young chemistry and strong detections of cyanopolyynes, constituting qualitative agreement with this dense core model. However, the maximum abundances of HC$_3$N and other cyanopolyynes calculated here fall some way short of those observed in TMC-1 core D ($X$[HC$_3$N]$\simeq 2 \times 10^{-8}$). This may be remedied to some extent by adopting different elemental abundances, but the most likely case is that, rather than evolving from the ``chemical equilibrium'' used here, TMC-1 core D has evolved directly from a largely atomic state, as modelled by \cite{hartquist01a}. The presence of even more free C/C$^+$ in our model at early times would improve agreement with observed cyanopolyyne levels (see section 2.1), for the first cycle. This could suggest that TMC-1 core D is a first cycle core at the hydrocarbon ``shoulder'' stage of its evolution, or an assembly of such cores. 

\cite{price03a} suggest that diffuse clouds may show signs of molecular enhancement as a result of dynamical processing. If diffuse clouds are indeed a transient phase of dynamically active gas, then our results would also predict significant molecular enhancements. The variety of chemistries displayed by diffuse clouds could be explained by lines of sight intersecting a number of cores in different stages of formation and dispersal. The large extents of molecular material in this model, subsequent to core dissipation into the diffuse background medium, could also explain the discrepancy in the ratio of CO (1-0) linewidth to H$_2$ column density, the so-called $X$-factor -- see e.g. \cite{magnani03a}. This conclusion is in accord with other work on this topic \cite[]{taylor93a,bell06a}.

\section*{Acknowledgments}
RTG thanks PPARC for a studentship, and the National Science Foundation for partial support. DAW acknowledges the support of a Leverhulme Trust Emeritus Fellowship.

\bibliographystyle{mn2e}
\bibliography{robbib_2006}

\label{lastpage}

\end{document}